\documentclass[aps,prl,twocolumn,groupedaddress,showpacs]{revtex4}
\usepackage{graphicx}
\usepackage[dvipsnames,usenames]{color}
\usepackage{color}
\usepackage{amsmath}
\usepackage{float}
\usepackage{amssymb}
\usepackage{amsthm}
\usepackage{mathrsfs}

\emergencystretch=\maxdimen
\begin{document}

\title{Reservoir computing and task performing through using high-$\beta$ lasers with delayed optical feedback}
\author{Tao Wang$^{1}$, Can Jiang$^{2}$, Qing~Fang$^{2}$, Xingxing Guo$^{1}$, Yahui Zhang$^{1}$, Chaoyuan Jin$^{3,4}$, Shuiying~Xiang$^{1}$}

\affiliation{$^1$State Key Laboratory of Integrated Service Networks, Xidian University, Xi’an 710071, China}
\affiliation{$^2$School of Electronics and Information, Hangzhou Dianzi University, Hangzhou, 310018, China}
\affiliation{$^3$College of Information Science and Electronic Engineering, Zhejiang University, Hangzhou 310007, China}
\affiliation{$^4$International Joint Innovation Center, Zhejiang University, Haining 314400, China}

\date{\today}

\begin{abstract}
Nonlinear photonic sources including semiconductor lasers have recently been utilized as ideal computation elements for information processing. They supply energy-efficient way and rich dynamics for classification and recognition tasks. In this work, we propose and numerically study the dynamics of complex photonic systems including high-$\beta$ laser element with delayed feedback and functional current modulation, and employ nonlinear laser dynamics of near-threshold region for the application in time-delayed reservoir computing. The results indicate a perfect (100$\%$) recognition accuracy for the pattern recognition task, and an accuracy of about 98$\%$ for the Mackey-Glass chaotic sequences prediction. Therefore, the system shows an improvement of performance with low-power consumption, in particular, the error rate is an order of magnitude smaller in comparison with previous works. Furthermore, by changing the DC pump, we are able to modify the amount of spontaneous emission photons of the system, this then allow us to explore how the laser noise impact the performance of the reservoir computing system. Through manipulating these variables, we show a deeper understanding on the proposed system, which is helpful for the practical applications of reservoir computing.
\end{abstract}


\maketitle 


\section{Introduction}
\label{section label}

With the developments of the Internet of Things (IoT) and big data, a growing number of exchanged information in networks leads to novel methods for information processing are highly desirable. Reservoir computing (RC), due to the ability of simplifying the implementation of recurrent neural networks~\cite{Qian2020, Jia2023, Zhang2023}, has been studied intensively in the past several decades~\cite{Maass2002, Jaeger2004, Sande2017, Guo2019}. The conventional RC architecture is a real network and typically featured by a large number of nonlinear nodes, we call this kind of RC as spatially distributed RC (SDRC)~\cite{Sande2017}. SDRC has been successfully demonstrated through using the arrays of semiconductor optical amplifiers (SOA)~\cite{Vandoorne2011}, coupled photonic emitters (CPE)~\cite{Takano2018} and silicon photonic chip (SPC)~\cite{Vatin2018}. However, these fully implemented reservoir architectures with a high number of nodes bring various technical challenges.

Alternatively, the innovative concept of a virtual network for RC using a single nonlinear node with a time delay feedback loop has been proposed recently~\cite{Appeltant2011}, and it is classified as a time-delay RC (TDRC)~\cite{Brunner2011, Ortin2017, Goldmann2020, Stelzer2020}. In comparison with SDRC, TDRC can further simplify the energy-efficient design and allows people to use a single photonic node in combination with a large number of virtual nodes spreading over different locations in a time-delay line~\cite{Vatin2019}. Finally, the number of nodes in the architecture can be modified by changing the length of delay lines. This method can effectively alleviate the physical node/connectivity restrictions of SDRC. Up to now, TDRC has been implemented in different schemes including electronics~\cite{Appeltant2011, Soriano2015}, optoelectronics~\cite{Larger2012, Hulser2022, Chen2019}, and optics~\cite{Li2022, Chemboa2020, Ashner2021}. These implementations have shown enhanced performance in benchmark tasks such as Mackey-Glass and Santa-Fe temporal series, whereas in real-world tasks e.g. optical channel equalization due to dispersion induced power fading, TDRCs have offered a power-efficient alternative to digital signal processing~\cite{Skontranis2022}.

The computation mechanism of TDRC originates from the nonlinear transform of the information to be processed onto a high-dimensional state space, assisted by the fading memory properties~\cite{Chen2020}. Through increasing the operating bandwidth of photonic systems and operating the hardware-efficient photonic topologies, people can target at high information processing speed. In this sense, photonic RCs are ideal candidates for analog processing of optical communication signals. Among different optical TDRC schemes, the system of semiconductor laser (SL) with time-delayed feedback has been used for generating broadband chaotic signals for applications such as chaotic encryption and physical random number generation, by biasing the SL well above the lasing threshold~\cite{Xu2017, Zhang2017, Estebanez2020}. Semiconductor lasers are compact and efficient light sources that offer several unique advantages including small size, fast response, low power consumption and high-efficiency~\cite{Wang2019, Wang2015a}. SL-based TDRC systems can exhibit enhanced dynamical bandwidth emission in presence of an additional optical injection signal~\cite{Torre2004, Nazhan2015}. 

In this work, we numerically investigate a photonic reservoir computing scheme based on a high-$\beta$ (refers to $\beta \geq 10^{-2}$~\cite{Deng2021}) semiconductor laser which is operated near threshold region. High-$\beta$ lasers possess small footprint and high efficiency for information processing~\cite{Deng2021}, favoring the development of RC-based neuromorphic visual systems towards high-speed, low-consumption, and easy-integration~\cite{Javanshir2022}. The main idea of this work is based on the delayed optical feedback and external carrier modulation, and consider the lasing transition region where the amplified spontaneous emission photons enable spiking dynamics in temporal~\cite{Wang2015, Wang2021}. We prove that the novel TDRC is not only efficient in dealing with several bench-marking tasks, including pattern recognition and Mackey-Glass chaotic sequences prediction, but the power consuming is also much lower than the previous ones by using low-$\beta$ lasers. In addition, we pay more attention on the impact of spontaneous emission noise on the performance of the TDRC system. The study could prove valuable in the development and optimization of TDRC system in the future.

\section{Computational concept and model}
\begin{figure}[!t]
\centering
  \includegraphics[width=3.3in,height=1.8in]{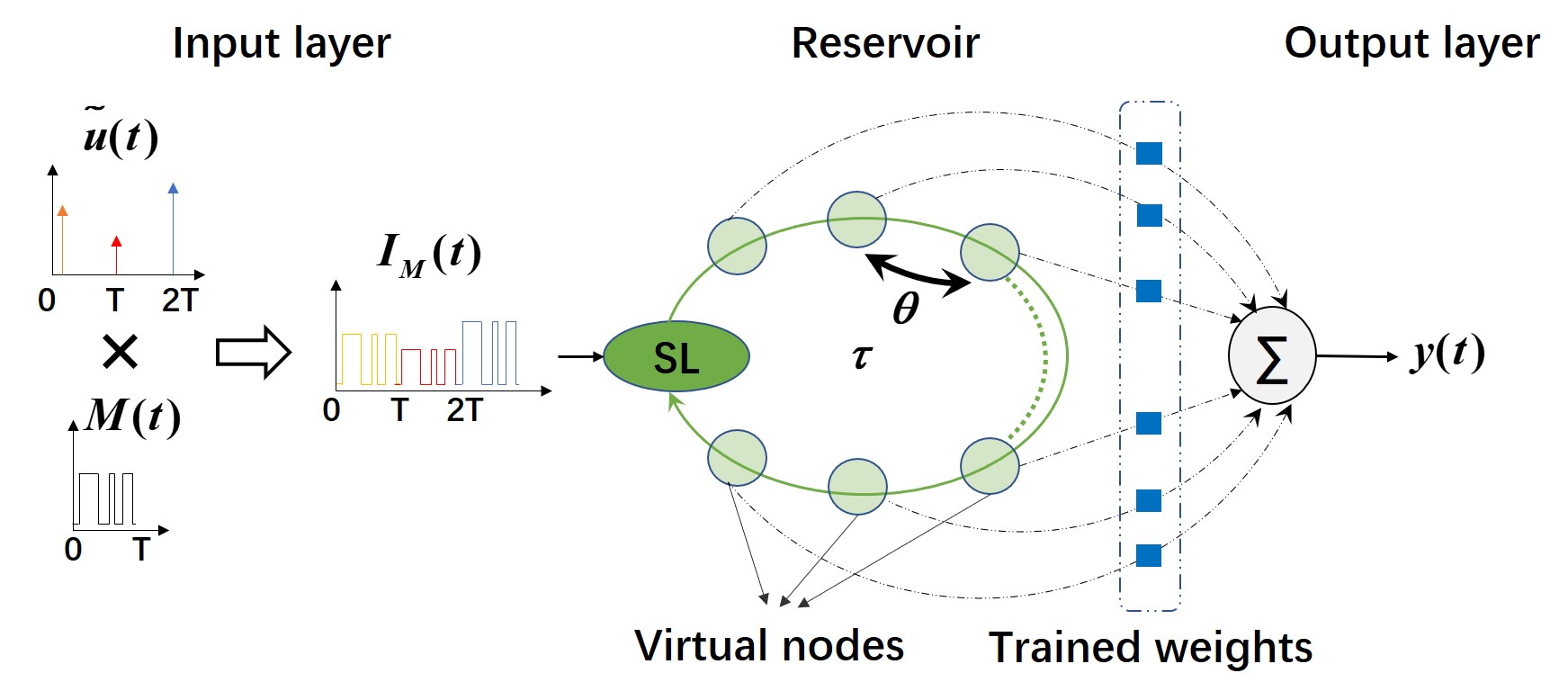}
  \caption{A photonic RC created with a high-$\beta$ SL subjects to time-delayed optical feedback, with $N$ virtual nodes defined by their equidistant time separation $\theta$.}
  \label{Setup}
\end{figure}

The scheme of the proposed TDRC based on a high-$\beta$ semiconductor laser (SL) is shown in Fig.~\ref{Setup}, where the RC is made up by three parts: the input layer, the hidden layer and the output layer. In the input layer, each of the discrete inputs $u(k)$ is maintained for an input time $T$ to produce a continuous signal $\widetilde{u}(t)= u(n)$ for $t \in [(k-1)T, T]$, where $T$ is the operation time of each sampling time, and it chosen close to the delay time $\tau$ in the feedback loop. Then input signal $\widetilde{u}(t)$ is subsequently multiplied by a step function $M(t)$ called input mask as well as a scaling factor ($\gamma$, it is a fixed random sequences) and the fluctuation range is ($-1$, $1$). This process is often called as masking, which enables the variable signal at different virtual nodes, and so that the information can be read out~\cite{Larger2012}, The resulted signal $I_M(t)$ is transformed into the input signal, and then coupled into the hidden layer through the modulator. 

In the hidden layer, which also refers to the reservoir part, through dividing the delay loop ($\tau$) into $N$ virtual nodes with spacing $\theta$, and $\tau = N\theta$. $\theta$ is the node separation which determines the time of the nonlinear node responds to the time-multiplexed input. In detail, if $\theta$ is larger than the intrinsic response time of the reservoir nodes, the latter response has enough time to settle down at a certain state~\cite{Brunner2022}. However, if $\theta$ is chosen small in comparison with the response time, the response of each nonlinear node will be always on a transient, each of them having a motion coupled to its close neighbors~\cite{Appeltant2011}. In addition, the amplitudes of the mask between neighboring $\theta$ are often selected randomly from a uniform distribution~\cite{Brunner2022}. Then, the time-multiplexed input signal with the imprinted mask drives the different virtual nonlinear nodes.

For the output layer, the output is interpreted as the state of the virtual network node, and the internal connection matrix, like the input matrix, is also a random sequences that is fixed and fluctuates at ($-1$, $1$). The output layer is a weighted and linear summation stage of the transient responses of different virtual nodes. Different from the input matrix and the internal connection matrix, the output matrix has to be trained. The linear least square method is used to minimize the mean square error between the target and predicted values, which can usually be determined by using computer offline training or programmable gate array on-line training. The optical signal is finally postprocessed in an offline procedure.

In order to properly simulate the proposed TDRC system based on a semiconductor microcavity laser with $\beta = 10^{-2}$, we adopt a newly developed \textit{Stochastic Simulator} (S-S)~\cite{Puccioni2015} by inserting the function modulation and feedback terms. S-S is specified as a semiclassical model, which can efficiently describe the dynamics for the lasers from macro- to microscale~\cite{Wang2021}. This model is unique in that it allows for the separate characterization of stimulated and spontaneous emission, which is different from the non-stochastic approaches. By separating different channel emission, the S-S model allows for more accurate analysis of laser dynamics. In addition, although the conventional rate equations possess the indubitable merit of providing a wealth of information about laser dynamics, they are failed to simulate the dynamics around threshold region, in particular for high-$\beta$ lasers, since people just introduce the averaged background noisy field into the lasing mode~\cite{Wang2022}. This incorrect description may also leads people to overlook the threshold region without any further consideration. In this work, we focus on the temporal dynamics of lasing transition region, which is often ignored when the laser is used for practical applications, due to the poor quality of coherence. However, for high-$\beta$ lasers, this region is very broad and show spike dynamics, possessing the potential in the application of reservoir computing.

In the terms of model, the detail recurrence relations between different physical processes have been well defined in~\cite{Puccioni2015} with the optical feedback added as in~\cite{Wang2021}. In addition to the parameters detailed in previous work ~\cite{Puccioni2015, Wang2021}, we set the spontaneous emission coupling factor $\beta$ of semiconductor laser to be $10^{-2}$, and ideally make the delay length to 0.6 m, which is corresponding to $\tau = 4$ ns time delay. For the Mackey-Glass chaotic sequences prediction, it is composed of $2.5P_{th} + 5 \times Mackey-Glass$ chaotic sequences ($P_{th}$ is the so called threshold pump~\cite{Rice1994}). 

To evaluate the performance of the proposed TDRC system, the tests of \textit{pattern recognition} and \textit{Mackey-Glass chaotic sequence prediction} have been carried out. The input dataset is composed of 7000 samples, of which 5000 samples are used for training and 2000 samples are for testing. The prediction performance of the system can usually be quantified by computing the normalized mean squared error (NMSE) between the target data $y_d$ and the RC system output $y_{out}$, which is defined by~\cite{Vatin2019}:

\begin{eqnarray}
{\rm NMSE} = \frac{1}{L} \frac{\sum_{n = 1}^L (y_d(n) - y_{out}(n))^2}{\sigma(y_d)}
\end{eqnarray}

\noindent Where $n$ is the subscript value of the data index, $L$ is the total data set, $y_{out}$ is the output value of the RC system, $y_d$ is the original data, and $\sigma (\cdot)$ represents the variance of the original signal.

\section{Performance characterization of TDRC system using high-$\beta$ lasers}
In this section, we will investigate the two different tasks of pattern recognition and Mackey-Glass chaotic sequences prediction respectively, and analyze the influences of scaling factor, number of virtual nodes, injection pump strength and feedback strength on the prediction performance, and then the memory capacity of the RC is also investigated. During the simulation, regression regularization parameters are used to optimize the performance of each task.

\subsection{Pattern recognition task}
\begin{figure}[!t]
\centering
  \includegraphics[width=3.4in,height=1.7in]{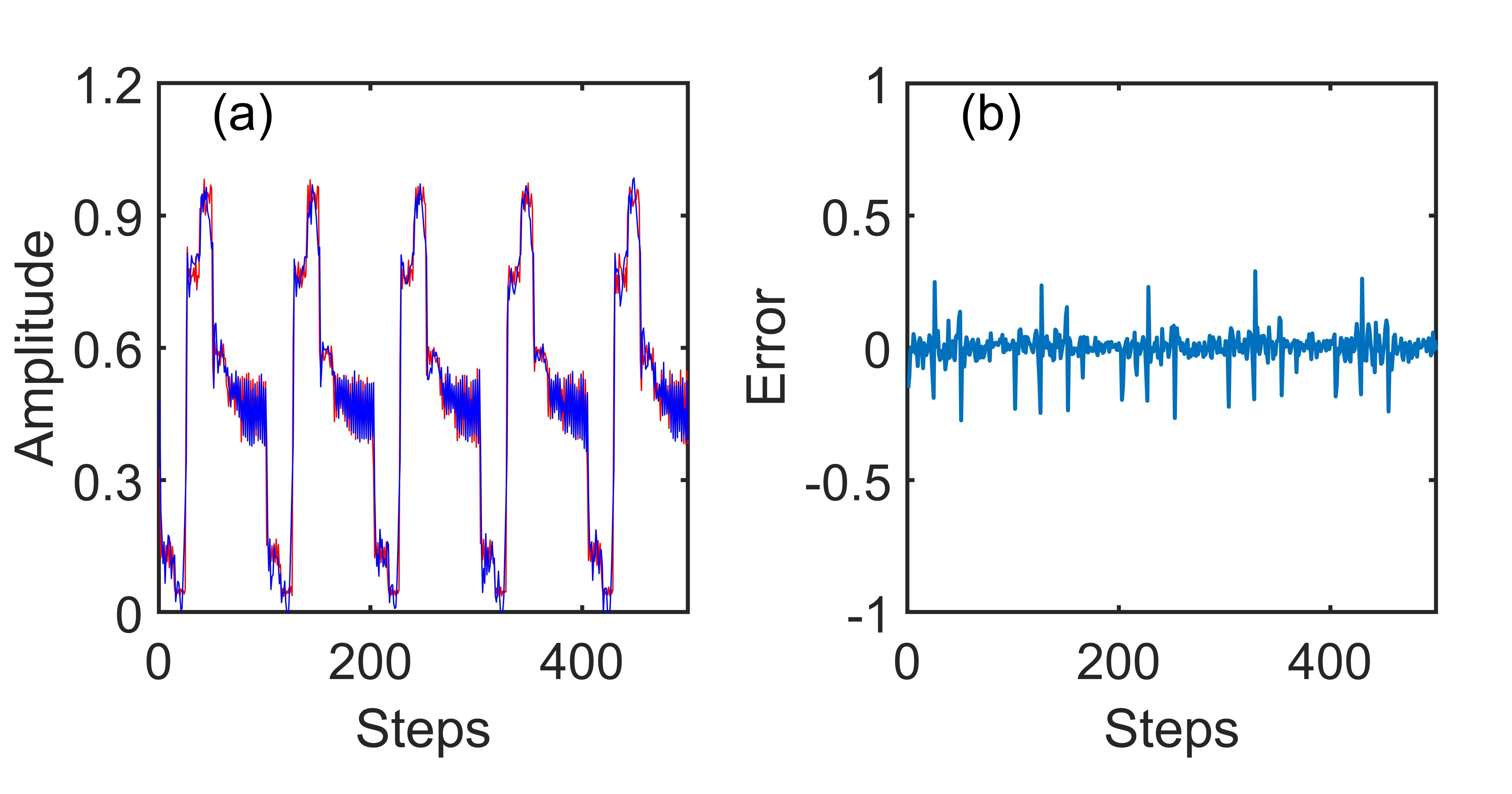}
  \caption{(a) Temporal waveforms of the original signal (red curve) and the predicted signal (blue curve) through TDRC system obtained under $\gamma = 0.81$, $N = 64$, $\eta = 0.2$ and $P_{DC} = 2.5P_{th}$; (b) Temporal errors after predicting by using TDRC.}
  \label{PatternR}
\end{figure}

The RC system is first used for a pattern recognition task, as shown in Fig.~\ref{PatternR}. The target of this task is re-generate the original input signal from the output layer. We set the scaling factor to be 0.51, ans make the virtual nodes $N = 64$. For the feedback, the strength is defined by $\eta = S_{inj}/ S_{out}$ ($S_{inj}$ indicates the photon number fed back into the laser, and $S_{out}$ is the photon number outcoupled from the cavity), is set to 0.2. It is worth to mention that the laser is biased at the position of $P_{DC} = 2.5P_{th}$, which is still in the lasing transition region. As mentioned before, the S-S model is more effective in describing the laser dynamics within this transition region, since it gives more attention on the dynamics of spontaneous emission noise. As shown by Fig.~\ref{PatternR}a, the temporal signal pattern indicated by the red curve is the original signal, and the blue curve is the predicted result by the TDRC system. After comparison of them statistically, we can observe that the predicted signal matches the original pattern well, the minimum NMSE value goes down to 0.0467. Fig.~\ref{PatternR}b shows the difference between the original and predicted signals. The smaller NMSE indicates the TDRC reaches a better prediction effect on the pattern recognition task in comparison with the previous work~\cite{Guo2020}.

\begin{figure}[!t]
\centering
  \includegraphics[width=3.4in]{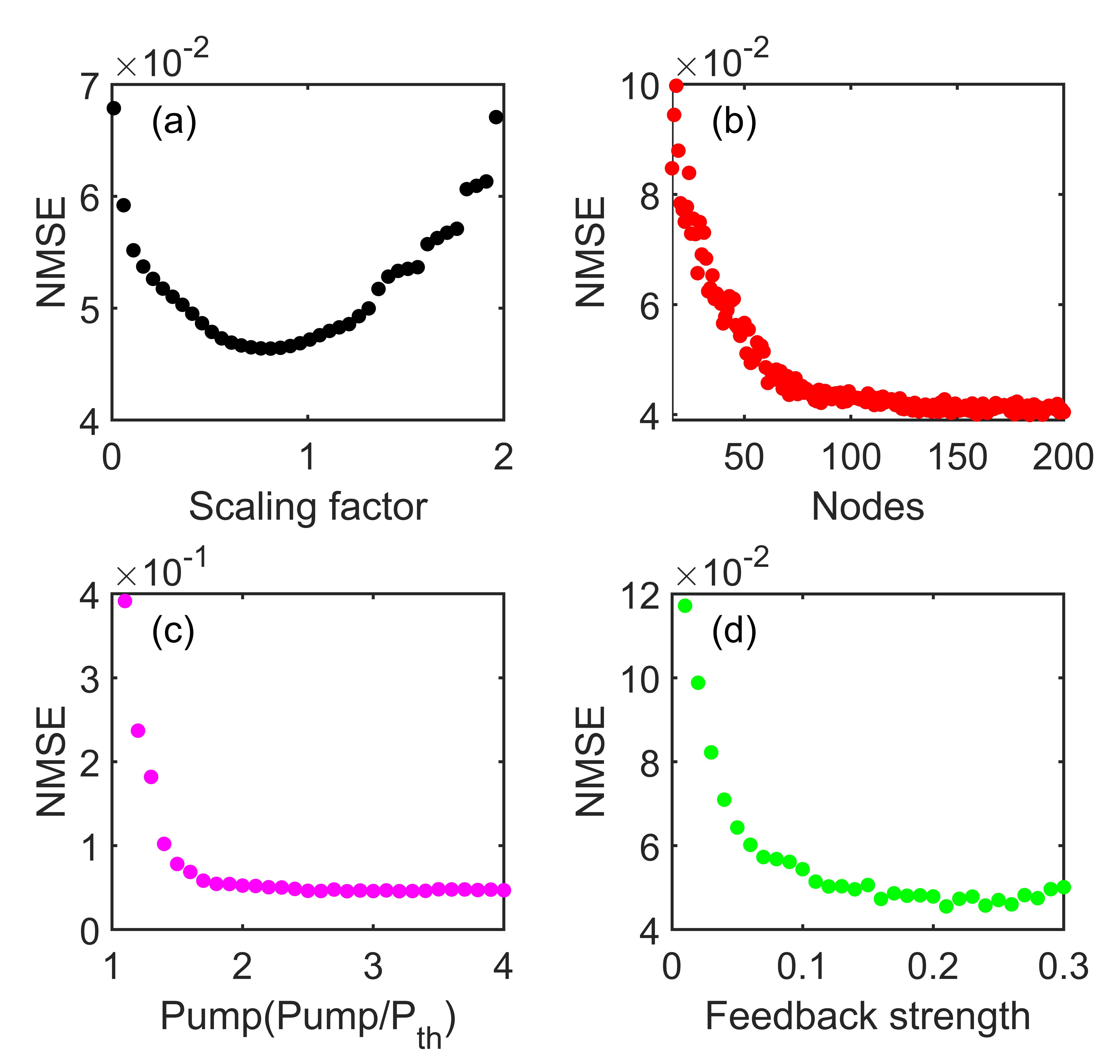}
  \caption{Influences of different parameters on the pattern recognition prediction: (a) scaling coefficient, (b) number of virtual node, (c) DC pump strength and (d) feedback strength.}
  \label{Parameters}
\end{figure}

In order to further explore the influences of different parameters on the predicted performance of the TDRC system, we carry out a series of investigations by changing the scaling factor, number of virtual nodes, injection strength and feedback strength, separately. Fig.~\ref{Parameters}a shows the variation of NMSE as the function of scaling factor $\gamma$, under the condition of $N = 64$, $\eta = 0.2$ and $P_{DC} = 2.5 P_{th}$. The function curve shows the value of NMSE first decreases with $\gamma$, reaching the minimum value at $\gamma = 0.81$, then goes up again. Thus, we can easily find the optimized prediction result is achieved when $\gamma = 0.81$, where the NMSE reaches to the minimum value, 0.046. Then, the change of NMSE with the number of virtual node is plotted by setting $\gamma = 0.81$. As shown in Fig.~\ref{Parameters}b, the NMSE values keep decreasing with the number of virtual node, but in different levels. Another slope is observed when $N > 64$, which indicates the decaying becomes very slow. After that, the effect of DC pump is also investigated by fixing $N = 64$. It is obvious that NMSE gives a substantial reduction when $P_{DC} < 1.7P_{tn}$ (Fig.~\ref{Parameters}c), which is mainly attributed by the reduction of spontaneous emission noise. But it almost keeps a constant when $P_{DC} > 1.7P_{tn}$. Fig.~\ref{Parameters}d is the influence of feedback strength on NMSE, and the function curve exhibits a similar line shape with the result of Fig.~\ref{Parameters}c. The value of NMSE cannot be further reduced when $\eta > 0.15$. 

\begin{figure}[!t]
\centering
  \includegraphics[width=1.65in, height=1.3in]{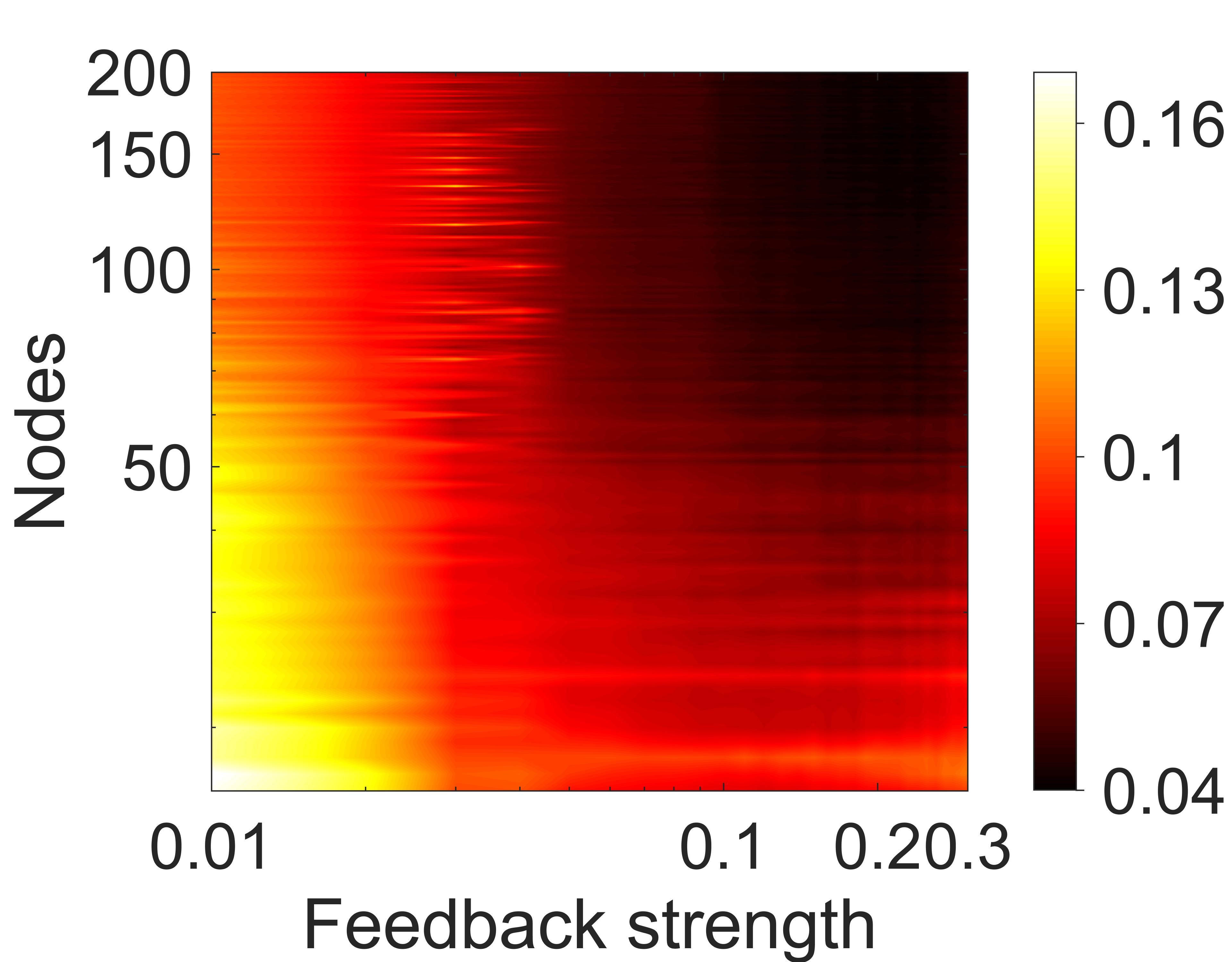}
  \includegraphics[width=1.65in, height=1.3in]{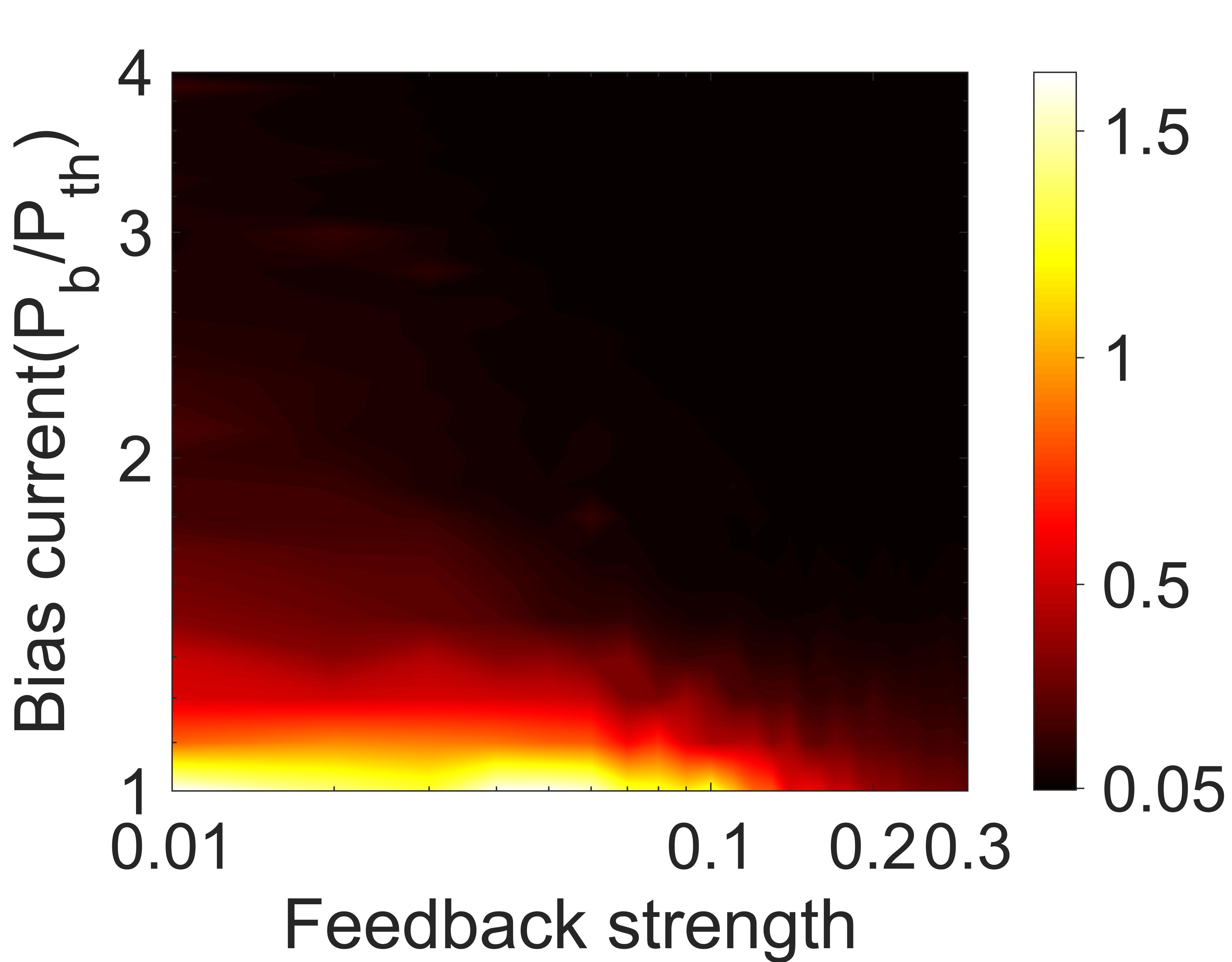}
  \caption{NMSE for the pattern recognition task as a function of feedback strength and virtual node number (left), as well as a function of feedback strength and DC pump (right) using the high-$\beta$-laser-based TDRC system.}
  \label{FandI}
\end{figure}

To further dynamically identify the optimized parameters, the simultaneous dependence of NMSE on the feedback strength and the virtual node number is investigated by plotting the distribution of NMSE. As shown in left panel of Fig.~\ref{FandI}, we find that NMSE achieves very small values with the order of $10^{-2}$ in the region of $\eta > 0.05$ and $N > 50$. For the conventional TDRCs using macroscopic lasers, it has been recognized that too small feedback level cannot supply sufficient memory for this task, and too large feedback can introduce system oscillation thereby degrades the consistency property~\cite{Brunner2013, Yue2019, Yue2021}. However, our results reveal that high-$\beta$ lasers enable TDRC still keeps good consistency even $\eta$ goes up to 0.30, showing a distinguish advantage. As for the virtual node, if the number is too small, there only a few sampled data which increases the fluctuations of the NMSE. However, too many virtue nodes will have no obvious effect on NMSE and even bring subtly adverse impact~\cite{Chen2019}.    


The right panel of Fig.~\ref{FandI} is the simultaneous dependence of TDRC performance on the DC pump and feedback strength. Interestingly, the results clearly show that NMSE can directly keep small values without the consideration of feedback strength if $P_{DC} > 1.3P_{th}$. In the opposite direction, the NMSE shows larger number since the large components of spontaneous noise, which degrades the performance of the system. To sum up, the feedback strength, the number of virtual nodes and the DC pump all impact the value of NMSE. However, if the reservoir works in the region with $P_{DC} > 1.3P_{th}$ and $N > 50$, the better prediction performance of the pattern recognition task can be obtained over a larger range of feedback strength. This remarkable feature enables the reservoir to be more flexible for the pattern recognition task.

\subsection{Mackey-Glass chaotic sequence prediction task}
Next, the TDRC system is also explored to predict the Mackey-Glass chaotic sequences, and the aim of this task is to perform single-point-prediction of chaotic data. In this simulation, the discretization time step is 0.3 ns. Additionally, the fixed scaling factor $\gamma = 0.81$, virtual nodes $N = 64$, feedback strength $\eta = 0.2$, and the DC pump is is still 2.5$P_{th}$. Then 7000 data points are generated by the high-$\beta$ laser, and 5000 points of the data for training and the left 2000 points for testing. 

\begin{figure}[!t]
\centering
  \includegraphics[width=3.4in,height=1.7in]{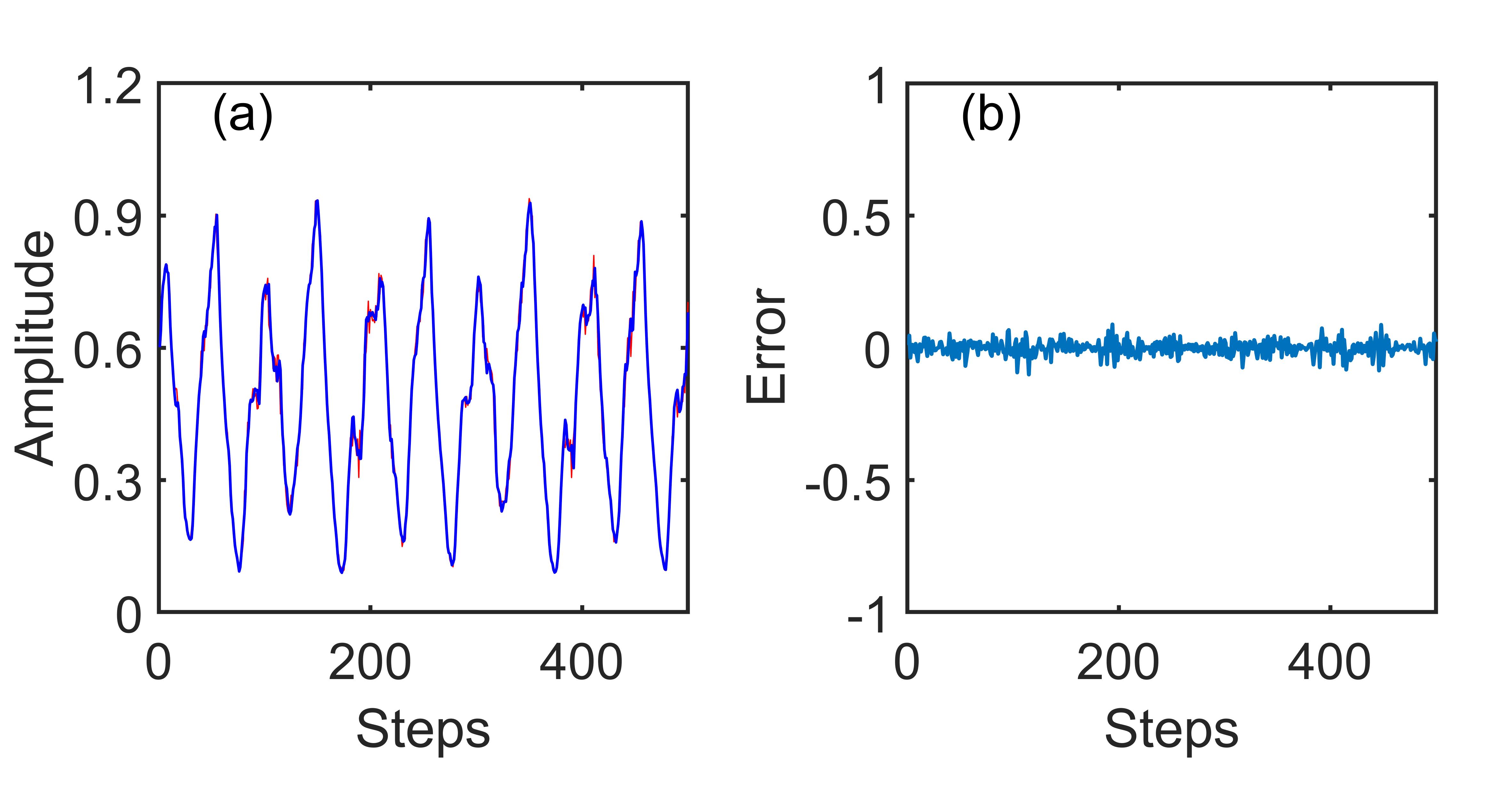}
  \caption{(a) Mackey-Glass sequence prediction time domain; (b) output matrix weights.}
  \label{MGSP}
\end{figure}

Fig.~\ref{MGSP}a shows the typical temporal waveforms for this task, and the predicted waveform (red curve) is obtained under the condition with $P_{DC} = 2.5P_{th}$, $\eta = 0.2$ and $N = 64$. we note that the red line almost coincides with the blue line, and there is only slight difference at the sharp positions. Therefore, the temporal deviations between the original and predicted signals are very small, and the errors are much smaller than the case of pattern recognition task, as shown in Fig.~\ref{MGSP}b.

\begin{figure}[!t]
\centering
  \includegraphics[width=3.4in]{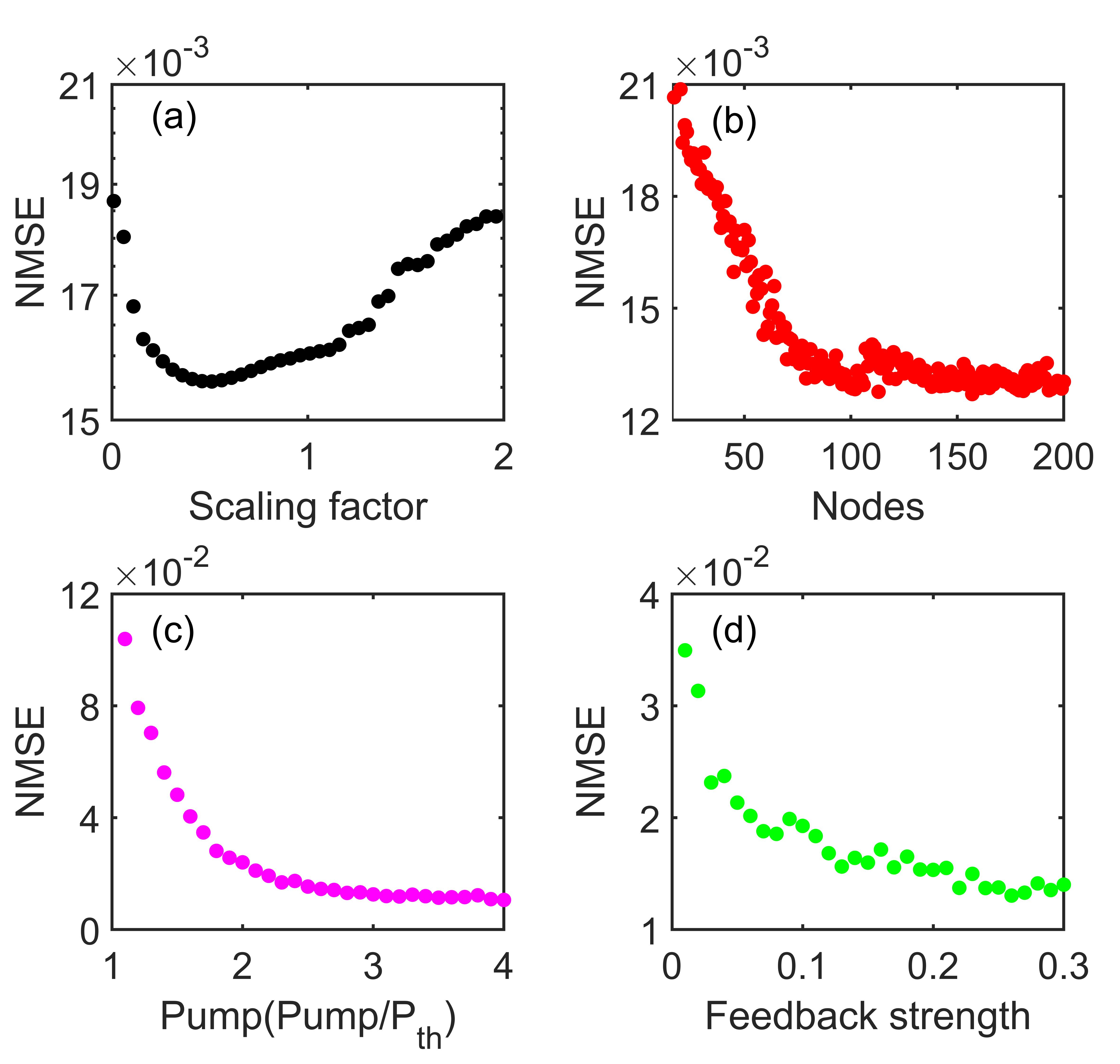}
  \caption{Effects of different parameters on the prediction of Mackey-Glass chaotic sequences: (a) scaling coefficient, (b) number of virtual nodes, (c) injection pump strength and (d) feedback strength.}
  \label{VNMSE}
\end{figure}

Same with the task of pattern recognition, we also investigate the effects of scaling factor, number of virtual nodes, injection pump strength and feedback strength on the prediction performance of Mackey-Glass chaotic sequences. Fig.~\ref{VNMSE}a is the effect of the scaling factor on NMSE, and the minimum NMSE of 0.015 is obtained when the scanning coefficient is 0.41. Fig.~\ref{VNMSE}b shows that a fast decaying of NMSE is observed within the region with $N < 80$, then a very slow decaying with small fluctuations is presented when $N > 80$, indicating a saturation of the performance improvement when $N > 80$. Fig.~\ref{VNMSE}b. Fig.~\ref{VNMSE}c and Fig.~\ref{VNMSE}d display the effects of DC pump and feedback strength on NMSE, respectively. In general, higher NMSE values ($\sim 10^{-2}$) are obtained at the beginning, then gradually decrease with DC pump and feedback strength, but the function curves display similar lineshapes. When $P_{DC} > 2.5P_{th}$ and $\eta > 0.2$, NMSE achieves the minimum value. When those parameters go up to even higher values, NMSE values almost keep a constant.

\begin{figure}[!t]
\centering
  \includegraphics[width=1.65in, height=1.3in]{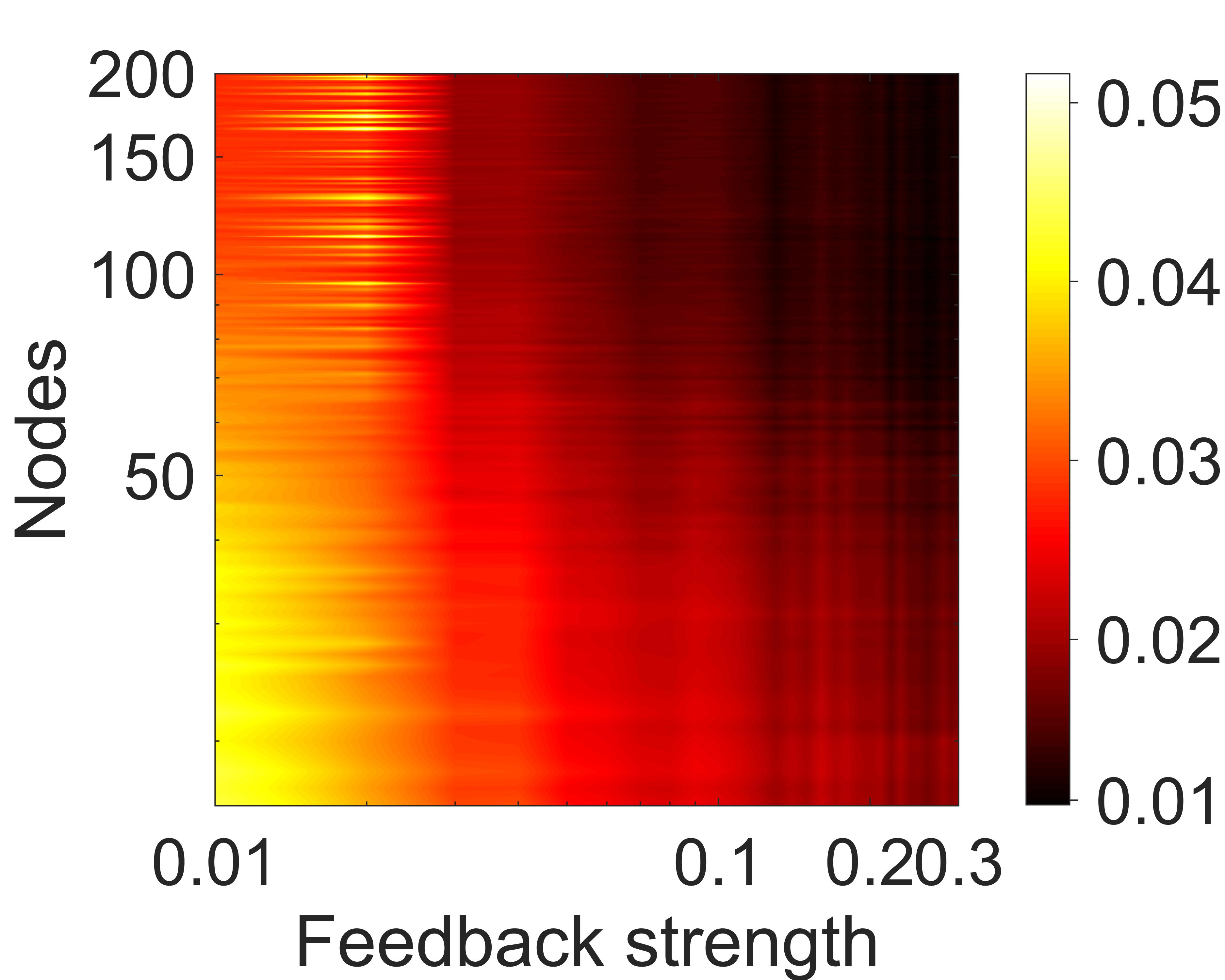}
  \includegraphics[width=1.65in, height=1.3in]{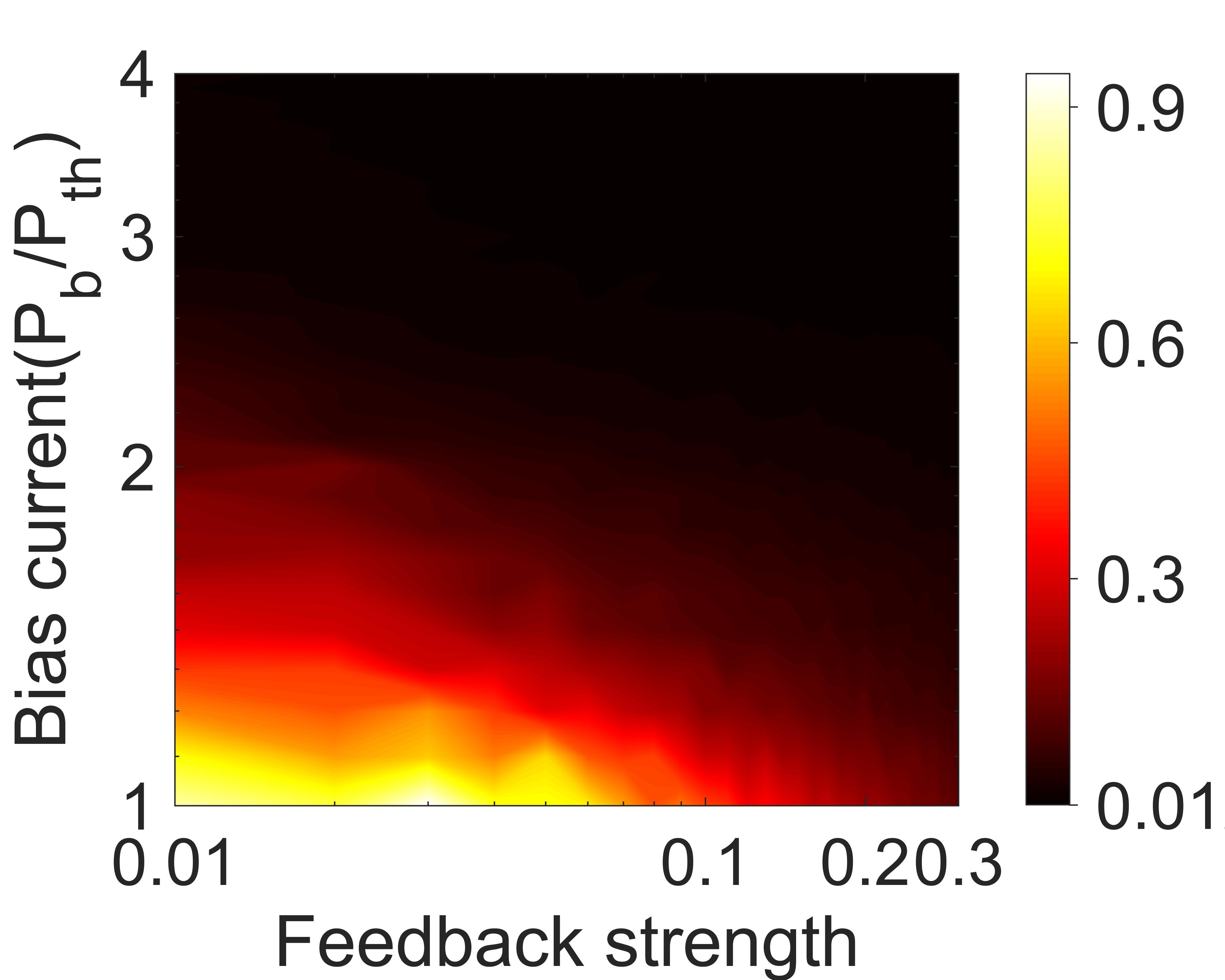}
  \caption{NMSE for the Mackey-Glass chaotic sequences prediction task as a function of feedback strength and virtue node number (left), also as a function of feedback strength and DC pump (right) using the high-$\beta$-laser-based TDRC system.}
  \label{IFNMSE}
\end{figure}

The left panel of Fig.~\ref{IFNMSE} shows the mapping diagram of NMSE distribution with the virtual node number and feedback strength, under the condition of $P_{DC} = 2.5P_{th}$. We note that the TDRC system has better performance than that of pattern recognition task, since the much smaller NMSE values. Closer observation indicates that the minimum NMSE locates in the region with $N > 60$ and $\eta > 0.2$. When the feedback strength is much lower, NMSE always keep higher values regardless how many virtue number it is. 


The right panel of Fig.~\ref{IFNMSE} is the mapping of NMSE distribution depending on the DC pump and feedback strength, under the condition of $N = 64$. When the laser is operated at $P_{DC} = P_{th}$, due to the large component of spontaneous emission noise, low level feedback doesn't give an obvious influence on the dynamics, thus, poor consistency of the system enables higher NMSE values. However, with the increase of DC pump, the coherence of laser output is improved. Then, for NMSE value with order of $10^{-2}$,  the required feedback strength is gradually decreased. When $P_{DC} > 2P_{th}$, a homogeneous NMSE distribution is achievable over a large range of feedback level, which is similar with the task of pattern recognition.

\section{memory capacity}
We also analyze the memory capacity (MC) property of the system, which is a key indicator for evaluation of TDRC's performance~\cite{Koster2021, Hulser2022b}. MC represents the property of the system to retain previously injected information, which is important when the prediction task requires past information, and quantifies the amount of information of past input signals that can be reproduced by the TDRC~\cite{Sugano2020}. For prediction tasks, it is especially important to have information from the past to accurately infer future values, but memory needs to disappear after a period of time so that responses are only influenced by the latest history, so it is necessary to discuss MC. MC is usually defined with a correlation function $m(i)$ as follows~\cite{Taylor1997}

\begin{eqnarray}
m(i) = \frac{Cov^2(y_i(n-i),O_i(n))}{\sigma^2(y_i(n))\sigma^2(O_i(n))}
\end{eqnarray}

\noindent where $y_i(n)$ is the injection signal, a pseudo-random sequence between the distributions (0, 1), $O_i(n)$ is the sequence of the RC response, $Cov$ is the covariance, and $\sigma^2()$ is the variance. MC is the sum of $m(i)$:

\begin{eqnarray}
MC = \sum_{i = 1}^\infty m(i)
\end{eqnarray}

\begin{figure}[!t]
\centering
  \includegraphics[width=3.4in]{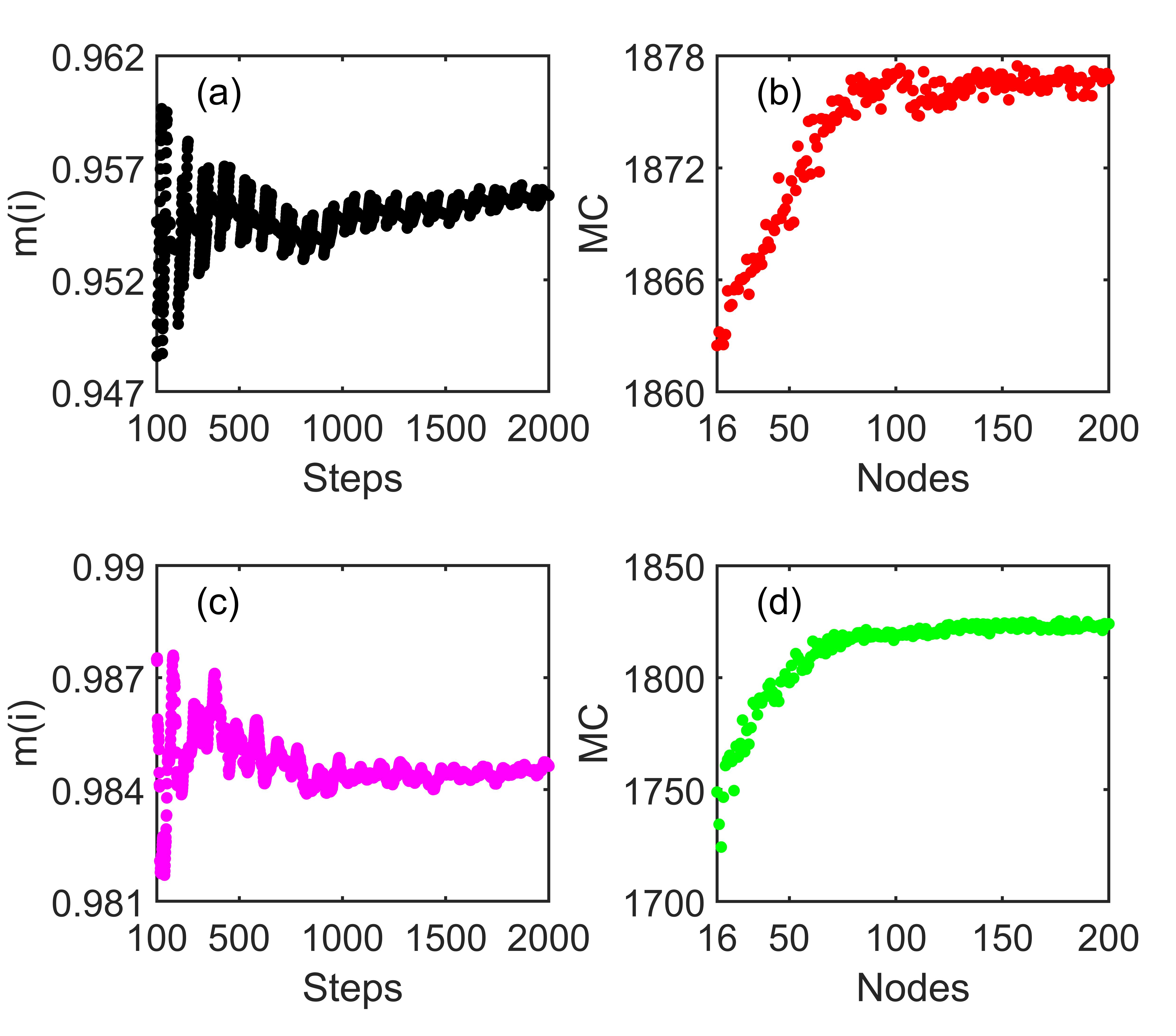}
  \caption{Effects of different parameters on the accuracy of pattern recognition (a-b) and prediction of Mackey-Glass chaotic sequences (c-d): for pattern recognition, (a) prediction steps, (b) number of virtual nodes; for the prediction of Mackey-Glass chaotic sequences, (c) prediction steps and (d) number of virtual nodes.}
  \label{PMMC}
\end{figure}

Fig.~\ref{PMMC} shows the calculated correlation functions and MC for the tasks of pattern recognition and Mackey-Glass chaotic sequences prediction. Fig.~\ref{PMMC}a is the correlation function curve $m(i)$ with different prediction steps for pattern recognition. It is clear to see that $m(i)$ keeps high value larger than 0.94, then it tends to be stable with the increase of the prediction step, although there are fluctuations. Fig.~\ref{PMMC}b is the variation of MC with the number of virtual nodes for the task of pattern recognition. Obviously, MC linearly increases with the number of virtual nodes. However, it starts to be saturated when $N > 100$. Thus, too many virtual nodes will result in a saturation on the MC, then degrade the prediction performance. Fig.~\ref{PMMC}c and Fig.~\ref{PMMC}d are the corresponding correlation function and MC for the task of Mackey-Glass chaotic sequences prediction. Similar with the pattern recognition task, $m(i)$ tend to be stable after a longer prediction step. The difference is that the values of $m(i)$ is higher than that for the case of pattern recognition, but the MC shows a shorter rang of virtual nodes before getting saturation.

\section{Influence of spontaneous emission noise}
Finally, in order to fully understand the influence of the spontaneous noise on the performance of TDRC system for different tasks, we conduct evaluations at $P_{DC} = 1.5P_{th}$, $2.0P_{th}$, and $2.5P_{th}$, respectively, and calculated the corresponding MC function curves with virtue node number. Typically, when we increase the DC pump, the laser emission becomes more coherent and the number of spontaneous photons decreases. By adjusting the DC pump, we can change the amount of spontaneous noise within the system and study how this affects the overall performance of the TDRC. 

\begin{figure}[!t]
\centering
  \includegraphics[width=3.4in,height=1.5in]{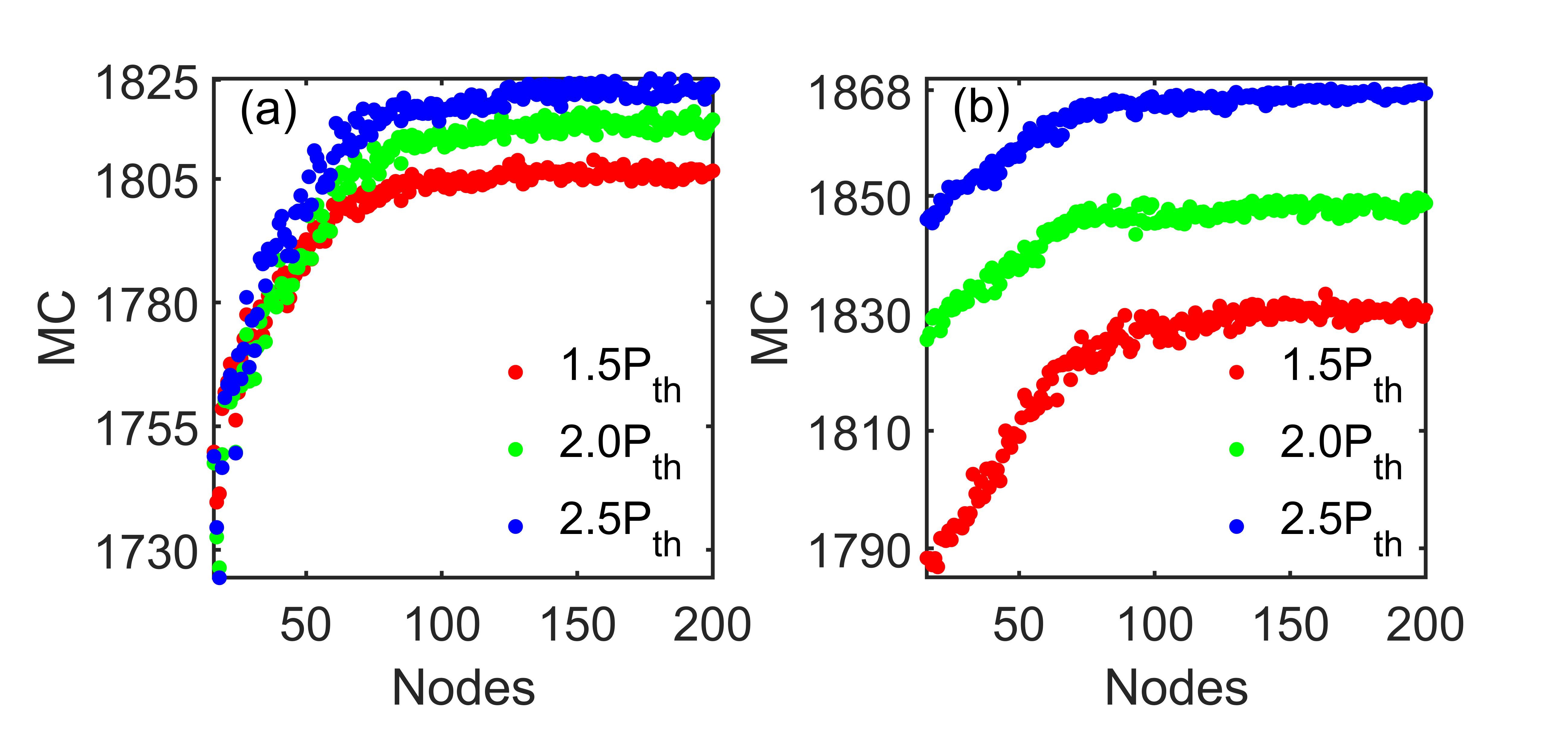}
  \caption{Calculated MC function curves for prediction of Mackey-Glass chaotic sequences (a) and pattern recognition tasks at $P_{DC} = 1.5P_{th}$, $2.0P_{th}$, and $2.5P_{th}$, respectively.}
  \label{Noiseinfluence}
\end{figure}

As shown in Fig.~\ref{Noiseinfluence}a, for the task of prediction of Mackey-Glass chaotic sequences, the variation DC pump has no obvious influence on the MC value in the region of $N \leq 50$. Therefore, the addition of noise doesn't affect the performance of TDRC when making predictions. However, there are noticeable deviations between the function curves when $N > 50$, and higher MC values are observed when using high DC pump. The results indicates that using fewer noises could potentially lead to a better performance. For the case of pattern recognition, as shown in Fig.~\ref{Noiseinfluence}b, the presence of spontaneous noise can have a significant impact on the MC value. It is clear to see that even before the saturation of MC, the negative effects of spontaneous noise on the performance of the TDRC are noticeable. This is particularly relevant since pattern recognition tasks tend to be more sensitive to noise, and the presence of spontaneous noise can make it more challenging to achieve accurate results. Therefore, it is essential to take spontaneous noise into account when designing and implementing TDRC systems for pattern recognition tasks, to ensure optimal accuracy and performance.

\section{Conclusions}
In summary, we have proposed a TDRC system based on the dynamics of threshold region of a high-$\beta$ semiconductor laser with delayed optical feedback, and carried out the numerical studies on the performance for the tasks of pattern recognition and Mackey-Glass chaotic sequence prediction. Since the unique feature of spiking dynamics in transition region, we found that semiconductor high-$\beta$ lasers display promising potentials for the TDRC system. Within the study, we pay special attention to the influence of the scaling factor, DC pump of the laser, virtual node number, and feedback strength on the prediction performance, and show it in the form of a mapping graph. The results show that the increase of those parameters within a certain range can significantly improve the prediction performance of the TDRC system. Therefore, properly adjusting those parameters is particularly important for TDRC to improve the prediction performance, thus, reducing the cost and the calculation time.

We have analyzed the MC properties for these two tasks, and the results indicate that with the increase of the number of virtual nodes, MCs are significantly improved. However, they tend to be stable after a certain range. This phenomenon explains that the continual increasing the number of virtual nodes cannot further improve the prediction performance of TDRC. Besides, we have also investigates how laser noise affects the performance of the TDRC system in terms of pattern recognition and Mackey-Glass chaotic sequence prediction. The results showed that the spontaneous noise had a significant negative impact on the overall performance of the TDRC system. Moreover, we found that the task of pattern recognition was more adversely affected by the spontaneous noise compared to chaotic sequence prediction. This indicates that the noise reduction should be considered to improve the performance of the TDRC system for applications that require high accuracy in pattern recognition. Our proposed TDRC system may hold promising potentials for the future on-chip reservoir computing. 

\section*{Acknowledgment}
The authors are grateful to the anonymous referees for constructive criticism and useful suggestions, which have improved the quality of the manuscript. They also thank to Prof. Gian Luca and Dr. Gian Piero Puccioni for fruitful discussions. This work is partially supported by several sources: National Key Research and Development Program of China (2021YFB2801900, 2021YFB2801901, 2021YFB2801902, 2021YFB2801904), National Natural Science Foundation of China (Grant No. 61804036), Zhejiang Province Commonweal Project (Grant No. LGJ20A040001), National Key R \& D Program Grant (Grant No. 2018YFE0120000), Zhejiang Provincial Key Research \& Development Project Grant (Grant No. 2019C04003).


\end{document}